\documentclass[english,debug,cite,overfull,info,nobm,showpacs]{epl}
\usepackage[T1]{fontenc}
\usepackage[latin1]{inputenc}
\usepackage{graphicx}
\usepackage{amssymb}

\newcommand{\noun}[1]{\textsc{#1}}
\newcommand{\boldsymbol}[1]{\mbox{\boldmath $#1$}}

\newcommand{\xib}{\boldsymbol {\xi}}
\newcommand{\etab}{\boldsymbol {\eta}}
\newcommand{\zetab}{\boldsymbol {\zeta}}
\newcommand{\al} { \tilde{\alpha} }

\newcommand{\exv}[1]{\langle#1\rangle}

\usepackage{babel}

\begin{document}

\title{Multi-electron giant dipole resonances of atoms in crossed electric and magnetic fields}
\shorttitle{Multi-electron giant dipole resonances in crossed fields}

\author{S.\ Z\"ollner\inst{1}\thanks{e-mail: \email{sascha@tc.pci.uni-heidelberg.de}} \and H.-D.\ Meyer\inst{1}\thanks{e-mail: \email{dieter@tc.pci.uni-heidelberg.de}} \and P.\ Schmelcher\inst{1,2}\thanks{Corresponding author; e-mail: \email{peter@tc.pci.uni-heidelberg.de}}}
\shortauthor{S.\ Z\"ollner \etal}
\institute{
	\inst{1} Theoretische Chemie, Physikalisch-Chemisches Institut, INF 229, Universit\"at Heidelberg, 69120 Heidelberg, Germany \\
	\inst{2} Physikalisches Institut, Philosophenweg 12, Universit\"at Heidelberg, 69120 Heidelberg, Germany
}

\maketitle

\begin{abstract}
Multi-electron giant dipole resonances of atoms in crossed electric and magnetic fields are investigated.
Stationary configurations corresponding to a highly symmetric arrangement of the electrons on a decentered
circle are derived, and a normal-mode stability analysis is performed. A classification of the various
modes, which are dominated either by the magnetic or Coulomb interactions, is provided. A six-dimensional
wave-packet dynamical study, based on the MCTDH approach, is accomplished for the two-electron resonances,
yielding in particular lifetimes of more than $0.1\,\mu$s for strong electric fields.
\end{abstract}

\section{Introduction}
Atoms exposed to strong external fields have proven to be a persistent source
of intriguing phenomena with major impact on a variety of other fields such as
the quantum dynamics of finite systems or Laser spectroscopy of Rydberg atoms 
\cite{Friedrich97,Schmelcher98,Gallagher94}. Focusing on crossed electric and
magnetic fields, new configurations of the hydrogen atom, the so-called giant
dipole states, were discovered in the nineties \cite{Dzyaloshinskii92,Baye92,Dippel94}.
These states represent a new form of microscopic matter where the electron
and proton are separated by large distances due to the presence of an electric 
field: There exists an outer potential well whose bound atomic states possess a huge
electric dipole moment. The atomic wave function then
resembles an oscillating barbell in contrast to the usual shell-like structure
of the electronic states in field-free space. In refs. \cite{Ackermann97}
the giant dipole states of Positronium have been investigated, thereby
arriving at the conclusion that the decentered positron-electron configuration 
is a quasistable state, i.e., the matter-antimatter system could be prevented
from annihilation for very large time scales up to several years. Recently, dipolar matter has also become of major interest in the context of ultracold atomic and molecular physics, such as dipolar quantum gases \cite{Griesmaier05} or ultracold molecular Rydberg states \cite{Greene00}.

In the present investigation we consider an $N-$electron atom (total mass $M$) where the
electrons ($m$) and the nucleus,
interacting via the Coulomb potential $V$, are subject to crossed
electric and magnetic fields. The translational invariance yielding the
conservation of the total canonical momentum in the absence of the fields 
is now lost, which manifests itself in the appearance of the coordinate-dependent
vector potential. For homogeneous fields the conserved
(total) pseudomomentum ${\bf{K}} = \sum_{i=0}^{N} {\bf{k}}_i$ with
${\bf{k}}_i = {\bf{p}}_i - e_i {\bf{A}}_i + e_i {\bf{B}} \times {\bf{r}}_i$,
whose components commute due to the neutrality of the atom, is a conserved
quantity associated with the center-of-mass (CM) motion of the atom \cite{Avron78,Johnson83,Herold81,Schmelcher94}.
It can be exploited to perform a so-called pseudoseparation of the CM motion.
The latter was originally done \cite{Avron78,Johnson83,Herold81,Schmelcher94}
for a fixed gauge. It is only recently that gauge-independent pseudoseparations
for one-electron \cite{Dippel94} and many-electron systems \cite{Schmelcher01}
have been performed. The latter provided a generalized potential for the 
electronic motion, possessing major impact on the prediction
of new structures and effects in crossed fields such as the decentered giant dipole
states \cite{Dippel94,Schmelcher01}. Our starting point is the pseudo-separated
Hamiltonian 
\begin{eqnarray}
{\cal{H}}  =  \frac{m}{2} \sum_{i=1}^N \dot{\bf{r}}_i^2 - \frac{1}{2M} \left( \sum_{i=1}^N m {\dot{\bf{r}}}_i \right)^2 \nonumber\\
\hspace*{-1cm} + \frac{1}{2M} \left({\bf{K}} -e {\bf{B}} \times \sum_i {\bf{r}}_i \right)^2 
-  e {\bf{E}} \cdot \sum_{i=1}^N {\bf{r}}_i + V,
\label{eq1}
\end{eqnarray}
where the first two terms and the last three terms represent the gauge-dependent
kinetic energy of the electrons relative to the nucleus 
(${\bf{\dot{r}}}_i = {\bf{\dot{r}}}_i (\{{\bf{p}}_j, {\bf{A}}({\bf{r}}_j)\}),\;i,j=1,\dots,N)$
and the generalized gauge-independent potential, respectively. The potential contains,
besides the Coulomb interaction $V$, Stark and diamagnetic interaction terms that
are responsible for the existence of an outer well and bound giant dipole states for
sufficiently strong (external and/or motional) electric fields in the one-electron
case.

Little is known, however, about multi-electron giant dipole resonances (GDR) which
are the subject of the present investigaton:
In Ref.\,\cite{Schmelcher01} a naturally limited analytical investigation of the decentered two-electron configuration
was performed. It provided evidence, but no final conclusion, on the existence of highly
symmetric decentered resonances. In the following we derive the electronic configurations corresponding to the decentered
$N-$electron giant dipole resonances. We classify the modes and investigate their stability.
Moreover, a numerically exact six-dimensional wave-packet dynamical investigation of two-electron
resonances will be performed, supplying valuable information on lower bounds of the lifetimes of the GDR.

\section{Stationary decentered configurations}
With the gauge-independent generalized potential at hand, let us search for decentered stationary points,
being candidates for configurations of GDR.
Inspecting the generalized potential (see Eq.\,\ref{eq1}), it is natural to introduce the 
electronic center of mass (ECM) ${\bf{R}}= \frac{1}{N} \sum_{i=1}^{N} {\bf{r}}_i$ 
as a new coordinate. In addition, we choose $N-1$ vectors relative to
the ECM, i.e., we decompose $\mathbf{r}_{i}=:\mathbf{R}+\mathbf{s}_{i},~i=1,\dots,N-1$.
Defining $Q=(\mathbf{R},\mathbf{s}_{1},\dots,\,\mathbf{s}_{N-1})^{T}$
we seek the stationary configuration
\begin{equation}
\frac{\partial\mathcal{V}}{\partial Q}(Q_{0})=0.\label{eq:stationarity}
\end{equation}
Employing the coordinate frame $\zetab=\mathbf{B}/B;\,\etab=\mathbf{K}'/K';\,\xib=\etab\times\zetab/|\etab\times\zetab|$
(where $\mathbf{K}'=\mathbf{K}+M\mathbf{E\times B}/B^{2}$, and we assume 
$\angle(\mathbf{K}',\mathbf{B})=90°$), one can show that there exist
solutions that fulfill $|\mathbf{r}_{i}^{(0)}|\equiv|\mathbf{R}_{0}+\mathbf{s}_{i}^{(0)}|=:r \;\forall i$
such that the ECM vector $\mathbf{R}=X\boldsymbol{\xi}+Y\boldsymbol{\eta}+Z\boldsymbol{\zeta}$
is aligned along the $\xib$ axis, $\mathbf{R}_{0}=(X_{0},0,0)^{T},$
where the decentering coordinate $X_{0}$ satisfies 
\begin{equation}
P_{r}(X_{0})\equiv-\frac{NeK'B}{M}+\frac{(NeB)^{2}}{M}X_{0}+NZe^{2}\frac{X_{0}}{r^{3}}=0.\label{eq:Pr(X)}
\end{equation}
Moreover, the relative coordinates are arranged on a
circle in the orthogonal complement, $\mathbf{s}_{i}=s\,(0,\cos\phi_{i},\sin\phi_{i})^{T}$.
For symmetry reasons, we demand that all relative coordinates
be distributed uniformly on that circle
\begin{equation}
\phi_{k}=\Phi_{N}k+\Delta,\quad\Phi_{N}\equiv 2\pi/N.\label{eq:ass-2}
\end{equation}
This procedure determines the decentered configuration only up to a
global rotation by an angle $\Delta\in[0,2\pi).$ However,
it allows us to fix the ratios of $|X_{0}|$ and $s,r$
\begin{eqnarray}
r & = & s\sqrt[3]{\frac{4N}{\sum_{k=1}^{N-1}(1/\sin\frac{\Phi k}{2})}}=:\alpha_{N}s=:\tilde{\alpha}_{N}|X_{0}|
\label{eq:alpha}
\end{eqnarray}
and thus to solve Eq. (\ref{eq:Pr(X)}). The solutions are 
\[X_{0}(K')=\frac{K'}{3NB}(2\cos\left({\displaystyle 
{\scriptstyle \frac{\theta+2\pi}{3}}}\right)-1)\]
where $\theta\equiv\arccos[2(K_{\mathrm{cr}}/K')^{3}-1]$, provided that $K'\ge K_{\mathrm{cr}}\equiv\frac{3N}{\al_N}\sqrt[3]{\frac{MB}{4}}$.
For any $B$ there is a critical value $K_{\mathrm{cr}}$ for
$K'$ at which the decentering sets in. Experimentally, the value of $K'$
can be controlled via the strength of the external electric field.
\begin{figure}
\begin{center}\vspace{-0.5cm}\includegraphics[%
  width=9cm]{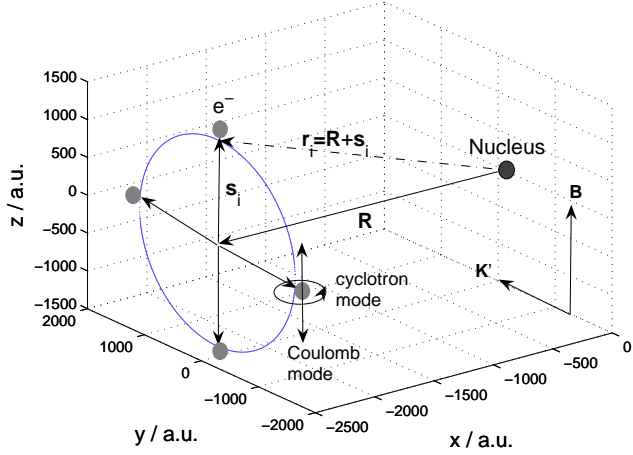}\vspace{-0.5cm}\end{center}

\caption{A giant-dipole configuration for $N=4$ electrons ($B=10^{-4},\, K=2K_{\mathrm{cr}}$).
The ECM $\mathbf{R}=(X,0,0)$ is decentered, with the relative vectors
$\mathbf{s}_{i}$ confined to a circle as indicated. Also shown is
the electronic vector relative to the nucleus $\mathbf{r}_{i}=\mathbf{R}+\mathbf{s}_{i}$.
Typical motions for the cyclotron and Coulomb mode
are indicated. \label{cap:giant-dipole-configuration}}
\end{figure}

To conclude, the stationary electronic vectors possess the orthogonal
decomposition $\mathbf{r}_{i}=X\xib+s(0,\cos\phi_{i},\sin\phi_{i})^{T}$
with the common decentering along $\xib$.
They are confined to a highly symmetric circular configuration perpendicular to $\xib$.
This circular configuration is determined up to an overall
rotation $\Delta$. A generic setup is sketched in Figure \ref{cap:giant-dipole-configuration}.
The extremal solution exists only if the effective pseudomomentum
$K'$ exceeds some critical value given above.
With increasing $K'$ the decentering of the ECM becomes more pronounced.

\section{Normal-mode analysis of $N$-electron giant dipole states}
We expect the decentered stationary configurations to be promising candidates for GDR
with certain lifetimes. Preceding
a numerical study, we first seek to obtain some insight into the local
stability of the extremal configurations. According to Ehrenfest's theorem, the expectation
values $\langle Q\rangle(t)$ of a harmonic system obey the corresponding
classical equations of motion. This suggests treating the problem
within a normal-mode analysis.

The equations of motion are obtained in terms of the displacements
$v(t):=Q(t)-Q_{0}$,\begin{equation}
\ddot{v}=\omega\cdot\dot{v}+A\cdot v.\label{eq:eom-for-v}\end{equation}
The antisymmetric $3N\times3N$ cyclotron matrix $\omega$
contains the Lorentz force and the harmonic matrix $A$
is built up essentially from the Hessian of the generalized potential.
The solution of this system of differential equations is given by
the span\begin{equation}
v(t)=\sum_{\rho=1}^{6N}(v_{\rho}e^{\gamma_{\rho}t})c_{\rho}\;(v_{\rho}\in\mathbb{C}^{3N};\,\gamma_{\rho},c_{\rho}\in\mathbb{C}),\label{eq:NM_sum}\end{equation}
 fulfilling the quadratic eigenvalue equation \begin{equation}
(\gamma_{\rho}^{2}I-\gamma_{\rho}\omega-A)v_{\rho}=0,\label{eq:QEP}\end{equation}
$I$ being the identity. Our stability analysis amounts to finding the complex
eigenvalues $\gamma_{\rho}=:\Gamma_{\rho}+i\Omega_{\rho}$
(whose imaginary parts are frequencies of a vibration about a stable
point, and whose real part corresponds to an instability), and the
eigenvectors $v_{\rho}$. The above quadratic eigenvalue problem
is solved by reducing it to the
standard linear eigenvalue problem 
\begin{equation}
\left(\begin{array}{cc}
0 & I\\
A & \omega\end{array}\right)u=\gamma u;\;u\equiv\left(\begin{array}{c}
v\\\dot{v} \end{array}\right).
\label{eq:QEP-linear}
\end{equation}%
%
Every eigenpair $(\gamma,v)$ of (\ref{eq:QEP}) with $\mathrm{Im}\gamma\neq0$
has a twin pair $(\gamma^{*},v^{*})$. Thus we 
treat the $6N$ modes as effectively $3N$ modes.
%

%
\begin{figure}
\begin{center}\includegraphics[%
  width=6.5cm,
  keepaspectratio,
  angle=-90]{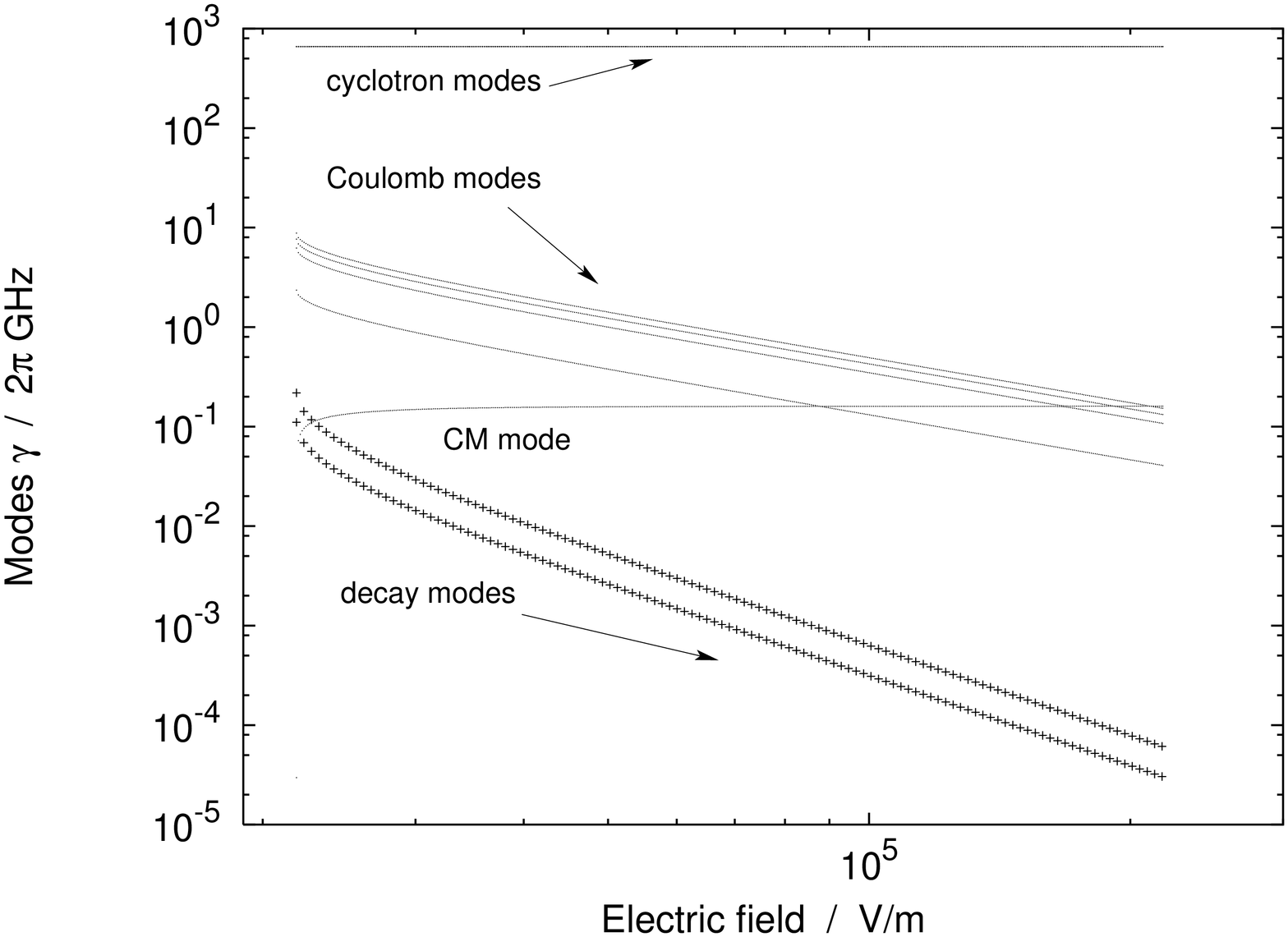}\end{center}

\caption{Eigenmodes $\{\gamma_{\rho}\}_{\rho}$ for $N=4$ electrons as a
function of the electric field $E\equiv BK/M$, shown for the range
$K/K_{\mathrm{cr}}\in[1,10]$ ($B=10^{-4}$). Imaginary parts appear as solid
lines, while points (+) are used for the real parts. -- The top horizontal
line represents four almost degenerate cyclotron modes. Below, the
four Coulomb modes fall off quickly and intersect the CM mode (the
nearly horizontal line about four orders below the cyclotron modes).
There are two residual decay modes, whose slope is twice that
of the Coulomb modes. \label{cap:NMA}}
\end{figure}
If one studies the behavior of the modes $\{\gamma_{\rho}(K;B)\}_{\rho}$
for different electron numbers $N$ in the vertical configuration $\Delta=-\pi/2$,
one observes that even though the patterns become increasingly rich
and involved with larger $N$, there is a clear distinction between the
character of the modes regarding their behavior as a function of $K$ (and $B$), their
order of magnitude and, more generally, their location in the complex
plane. Based on our analysis, including the associated eigenvectors,
we arrive at the following classification of the $3N$ eigenmodes:

\noindent
{$\bullet$} $N$ so-called cyclotron modes corresponding to the cyclotron
motion of the effective electronic particles. Their values are exclusively of 
the order of the cyclotron frequencies, $\gamma=i\Omega\sim i\frac{|e|B}{m}$,
i.e., almost independent of $K$.

\noindent
{$\bullet$} $N$ Coulomb modes corresponding to the inter-Coulombic
motion, with dominating contributions from the harmonic matrix $A$. They
fall off quickly with $K$, since the decentering $X_{0}(K)$ increases,
and the Coulomb interaction weakens. As opposed
to the cyclotron modes, the Coulomb motion takes places predominantly
parallel to $\mathbf{B}$.

\noindent
{$\bullet$} 1 CM mode, roughly reflecting the cyclotron motion of the CM,
$\Omega\sim\frac{NeB}{M}$. 

\noindent
{$\bullet$} 1 zero mode $(\gamma_{\rho}=0)$ stemming from the rotational
invariance of the saddle point with respect to the circular configuration
(see Eq. \ref{eq:ass-2})

\noindent
{$\bullet$} $N-2$ modes termed decay modes in recognition of the fact
that they are predominantly real. They are neither directly
related to the cyclotron motion nor to the spectrum of $A$, and their
slope is twice as steep as that of the Coulomb modes. Their absence
for the case $N=2$ is essentially why the two-electron system is
locally stable.

A generic example for the spectrum is given in Fig. \ref{cap:NMA} for the case of
$N=4$ electrons. There are a few subtleties that go beyond the categorization
suggested above. Without going into the details, we mention that there
are certain interactions among different mode types. Their principal
causes are crossings between the CM mode and the decay
modes (resulting in some striking deformations of the usual line pattern),
and avoided crossings of the CM mode with at least some of
the Coulomb modes. This may be taken as a hint for the different symmetry
relations among the Coulombic and the decay modes.

Let us remark on the influence of the global rotations $\Delta$.
While the stationary character is not affected by a common
rotation of the relative coordinates $\mathbf{s}_{i}$ by an angle
$\Delta$ on the circle, the dynamics differs. In order to see this
dependence, we inspect the modes as a function of the rotation angle
$\{\gamma_{\rho}(\Delta)\}$. Apart from an obvious symmetry---note
that rotating by $\Phi_{N}=2\pi/N$ gives an indistinguishable setup---we
found that the Coulomb modes and the decay modes show a pronounced
periodic change. For the case $N=2$, we encounter an exceptional behavior
whenever $\Delta$ comes close to $0\,\mathrm{mod}\,\pi$. One of the Coulomb modes
then tends to zero, along the way turning real. In this respect,
the local stability indicated above is applicable only outside the
singular horizontal configuration $\Delta=0$.

\section{Wave-packet dynamical study} 
We now turn to a numerical study
of the two-electron system ($N=2$). We emphasize
that a six-dimensional resonance study is at the frontier of what
is currently possible and requires a careful choice of the computational
approach. This applies especially in view of the fact that our system
is governed by dramatically different time scales (see below). Therefore
we adopted the Multi-Configuration Time-Dependent Hartree \noun{(mctdh)}
method \cite{Beck00,Worth00}, a wave-packet propagation scheme known for its outstanding efficiency
in high-dimensional settings. Its basic idea is to solve the time-dependent
Schrödinger equation by expanding the wave function in a moderately sized
time-dependent basis related to Hartree products. We stress that this
approach is designed for distinguishable particles. Applying it to
a fermionic system like ours finds its sole justification in the fact
that the spatial separation between the electrons is so large that
they are virtually distinguishable.

Let us first point out how the computational method can be applied
to our problem. There are essentially two conflicting types of motion 
and corresponding scales: perpendicular to the magnetic field,
we have the magnetic length $R_{B}\sim10^{2}\,\mathrm{a.u.}$ (for
the strong laboratory field strength $B = 10^{-4}\,\mathrm{a.u.}$ we consider). For the
Coulomb motion parallel to ${\bf{B}}$---expected to take place approximately
in a harmonic potential---we can use the usual oscillator lengths
$z_{0}\equiv1/\sqrt{m\omega_{z}}$ as an estimate. These are roughly
on the order of $10^{3}$ a.u. but increase with $K$ (see above).
Analogously, the anticipated time scales are $T_{B}=\frac{2\pi}{\omega_{B}}\stackrel{B=10^{-4}}{\simeq}1.5\,\mathrm{ps}$
for the cyclotronic motion and $1.5\mathrm{ns}-15\mathrm{ns}$ for
the Coulomb motion.

Within our numerical study we focus on the investigation of the stability of
the giant dipole states by using the propagation
of harmonic-oscillator wave packets initially localized at the extremal position.
Relaxation techniques have been applied in order to improve the intially chosen wave packet.
Moreover, following the evolution of wave packets with an initial
displacement from the extremum, we tested the robustness of the resonances.
For comparison, we also examine the decentered eigenvectors.
The chosen parameter set is $K/K_{\mathrm{cr}}\in\{1.1,\,2.0,\,10.0\}$
(we drop the prime here and in the following), which accounts for the cases of just above
threshold, the medium range and the very large $K$ regime. For
simplicity, we first focus on the vertical configuration $\Delta=-\pi/2$.
The propagation times are chosen in the regime of $50-100\,\mathrm{ns}$;
this includes many periods of the Coulomb modes and some $10,000$
periods of the rapid cyclotron motion !

Very close to the critical point, e.g. $K=1.1K_{\mathrm{cr}}$, the observed motion
is unstable for several degrees of freedom, specifically for
those belonging to the relative vector $\mathbf{s}_{\perp}=(x,y)^T$ perpendicular
to the magnetic field, on a time scale of $10^{3}\,\mathrm{ps}$.
Other modes are affected too via couplings.
This is a general fact, but for $K=1.1K_{\mathrm{cr}}$ it is very pronounced,
whereas it is suppressed for the time evolution in case of larger
values of $K$ due to the weaker Coulomb modes (see Fig. \ref{cap:NMA}).

\begin{figure}
\begin{center}\includegraphics[%
  width=7.5cm,
  keepaspectratio]{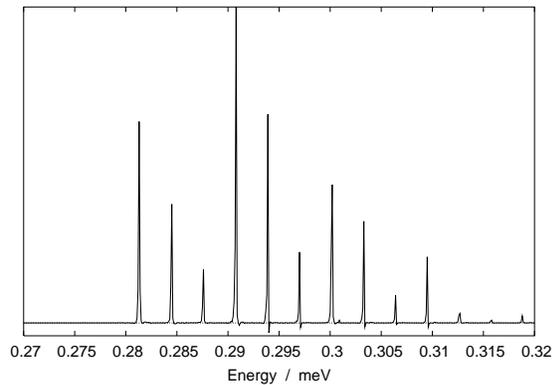}\vspace{-0.5cm}\end{center}

\caption{Spectrum for $K/K_{\mathrm{cr}}=2$ for the excited initial state (displaced
in $Z,z$).\label{cap:K2.0_spec}}
\end{figure}

For $K$ being twice its critical value, $K=2K_{\mathrm{cr}}$, the vertical
configuration is virtually stable on a scale of $T\sim10^{4}\,\mathrm{ps}$.
As opposed to the case $K/K_{\mathrm{cr}}=1.1$, the instability is almost
exclusively due to the relative motion in $y$. However, the response
of the system upon displacing the initial wave packet in $Z$
and $z$ by $2,000\,\mathrm{a.u.}$ unveiled that $x$ is rendered
rather unstable. Hence the resonance is expected to be less robust,
although its lifetime is not significantly reduced altogether. To
illustrate the resonance character of the system, Fig. \ref{cap:K2.0_spec}
gives an impression of the excitation spectrum, obtained via
Fourier transformation of the auto-correlation function $c(t):=\langle\Psi_{0}\mid e^{-iHt}\Psi_{0}\rangle$.
 The `displaced state' with $\exv{z}_{0}=\exv{Z}_{0}=2,000\,\mathrm{a.u.}$
produces a rather interesting excitation spectrum. The equidistant
spacing of the peaks can be interpreted as a signature of harmonicity
for both excited degrees of freedom ($Z,\, z$). 

For $K=10K_{\mathrm{cr}}$, we find the system to be practically stable on
the time scale $T=10^{5}\,\mathrm{ps}$. The initial
wave packet experiences only tiny deformations with respect to all degrees of freedom
but $z$, showing some oscillatory behavior. However, those are
still marginal compared to the spatial extension of the decentered
state. As a result of our wave-packet dynamical study, we can conclude
that the lifetimes of the GDR for sufficiently large $K$ are
beyond $0.1 \mu$s.

To complete the discussion, let us touch on the effects of different
circular configurations $\Delta$ of the GDR.
As examples, we investigated both the supposedly unstable {}`horizontal
configuration' $(\Delta=0)$ and a {}`diagonal configuration' $(\Delta=-\pi/4)$,
corresponding to different settings of the extremal relative coordinates
$\mathbf{s}^{(0)}$. To sum up our findings, the horizontal configuration
indeed adds an instability, which is discernible even for very high
$K$, if less distinct. The diagonal configuration was partly unstable
on a timescale comparable to that of the vertical case, $T\sim10^{4}\,\mathrm{ps}$. 

\section{Conclusion and outlook}
Our investigation shows that atoms in crossed fields exhibit 
multi-electron giant dipole states with extraordinary lifetimes.
These resonances constitute highly symmetric and exotic states of matter, where
the elecrons are strongly correlated and can, for laboratory field strengths,
be separated from the nucleus (or positively charged core)
by many thousand Bohr radii. An experimental preparation of the giant
dipole states might employ the scheme suggested for single electrons
\cite{Averbukh99}. The latter is based
on the preparation of Rydberg states via Laser excitation, followed by
a sequence of electric-field switches that carry the excited electrons
to the decentered configuration. Preparing different initial Rydberg states, like circular ones, enhances the variety of accessible decentered states.

\end{document}